\documentclass[aps,prl,twocolumn]{revtex4}
\input epsf
\begin{document}
\title{Dewetting of a solid monolayer
}
\author{Olivier Pierre-Louis$^1$, Anna Chame$^2$, Yukio Saito$^3$}
\affiliation{$^1$CNRS/Rudolf Peierls Centre for Theoretical Physics,
1 Keble Road, Oxford OX1 3NP, UK.\\
$^2$ Universidade Federal Fluminense,Avenida Litor\^anea s/n, 24210-340 Niter\'oi RJ, Brazil\\
$^3$ Keio University, 3-14-1, Hiyoshi, Kohoku-ku, Yokohama, Kanagawa, 223-8522 Japan
}
\date{\today}

\begin{abstract}
We report on the dewetting of a monolayer on a solid substrate,
where mass transport occurs via surface diffusion. 
For a wide range of parameters, 
a labyrinthine pattern of bilayer islands is formed.
An irreversible regime and a thermodynamic
regime are identified. In both regimes, the velocity
of a dewetting front, the wavelength of the bilayer island 
pattern, and the rate of nucleation of dewetted zones
are obtained. We also point out the existence of a scaling behavior,
which is analyzed by means of a geometrical model.

\end{abstract}


\maketitle

Liquid films,
once spread on a substrate, may
break-up into droplets to lower the surface
energy. Such a process 
is called dewetting. 
As for liquids, thin solid films 
may break-up into droplets.
However two main differences may be pointed out.
Firstly, solids exhibit strong surface anisotropy
whereas liquids are usually isotropic. 
Secondly, mass transport mainly occurs via surface
diffusion on solids at small scales, while
it is mediated by hydrodynamics in liquids.

Dewetting of solid layers with sub-micron thicknesses 
was observed in
recent experimental studies \cite{Jiran,Yang2005,Krause}. 
Spontaneous breakup of the film
into dots can be analyzed
within the frame of continuum models including
an effective wetting potential, with
surface energy\cite{Golovin}, and elastic effects\cite{Golovin2}.
Moreover, the nonlinear dynamics of the edges 
of these layers \cite{Yang2005,wong,barbe}
may also lead to
the periodic formation of  holes behind the dewetting rim.

For even  thinner films, such as
1nm thick Ag on Si\cite{Thuermer},
one expects the discreteness of the underlying
crystalline lattice to come to the fore.
In order to investigate these effects,
we study the dewetting of the thinnest possible layer:
a monolayer. 
In order to focus on the basic processes,
we discard effects related to
substrate roughness, elastic interactions, or alloying.
We focus on the case
where dewetting occurs via the nucleation of 
holes, subsequently invading 
the whole film. 
This occurs in a well defined
temperature window: if the temperature
were too low the surface would be frozen;
if it were too high  --above the roughening
transition-- the
film would be unstable and would break up
into a microscopically disordered pattern.

We show that monolayer dewetting 
proceeds differently from thicker layers dewetting.
As shown on Fig.1, 
monolayers initially lead to a labyrinthine
pattern of bilayer islands,
which then slowly thicken into 3-layer, and then 4-layers islands, etc. 
Two different regimes for monolayer dewetting,
henceforth denoted as regimes I and II, are analyzed.
While both regimes exhibit the same temporal scaling
behavior, their microscopic dynamics
is qualitatively different.

We employ Kinetic Monte Carlo (KMC)
simulations in order to mimic experiments
with a minimum number of ingredients. We use a
Solid on Solid model on a 2D square lattice,
with lattice unit $a$,
and local height $h\geq 0$. The substrate
surface, at $h=0$, is perfectly flat and frozen.
Epilayer atoms hop to nearest neighbor sites with the rates 
$r_n$ when $h=1$, and
$\nu_n$ when $h>1$, with
\begin{eqnarray}
r_n = \nu_0\,{\rm e}^{-nJ/T+E_S/T};
\hspace{1 cm}
\nu_n = \nu_0\,{\rm e}^{-nJ/T}
\end{eqnarray}
where $\nu_0$ is a constant frequency,
$T$ is the temperature (in units with $k_B=1$),
$n$ is the number of in-plane nearest neighbors,
$J$ is the bond energy, and $E_S$ is the adsorbate-substrate
interface energy.
When $E_S/J$ is large, 
the adsorbate minimizes the total energy by creating
high islands with steep sides
\footnote{At low temperatures, the equilibrium island 
shape has a height $h_{eq}$ a square base of lateral size $L_{eq}$,
with $h_{eq}/L_{eq}=E_S/J$.}. 
When $E_S\rightarrow 0$, the adsorbate wets
the surface. We use
periodic boundary conditions, and lattice sizes ranging
from $200\times300$ to $10^3\times10^3$. 
The initial condition
is either: (IC1) a complete monolayer with $h=1$ everywhere, or (IC2) 
a monolayer with uncovered substrate ($h=0$) in a straight stripe
(of width 10a).
In all cases, the dewetting front invades
the whole substrate, leaving a labyrinth of bilayer 
islands with a width $\lambda$  in a denuded zone behind.
In IC2, with small enough $E_S$ and $T$,
the dewetting front starts from the existing monatomic
step and invades the monolayer at constant velocity
$V$, while in the other cases, holes 
are nucleated and closed fronts expand in the
film.

\begin{figure}
\centerline{\epsfysize=7cm\epsfbox{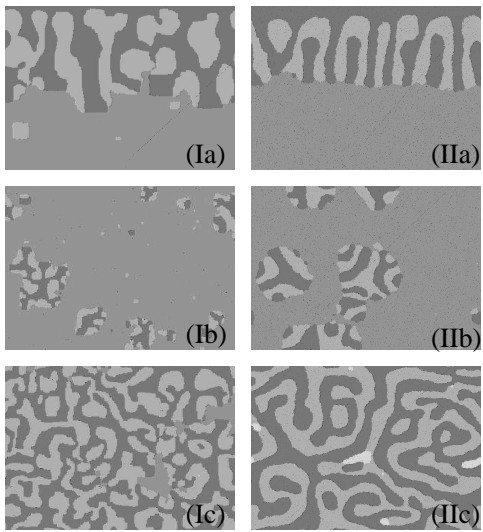}}
\caption{
KMC simulations of monolayer dewetting in regimes I and II.
Lighter means higher. (a): IC2,
(b): IC1, early dynamics; (c): IC1 late dynamics.
$E_S/J=0.75$, $J/T=5$ in (Ia,b,c).
$E_S/J=0.1$, $J/T=2.5$ in (IIa), and
$E_S/J=0.12$, $J/T=0.5$ in (IIb,c).
}
\label{fig1}
\end{figure}

Regime I  is observed for low $T/J$, and moderate $E_S/J$. 
On a mono-atomic
step which separates two bilayer islands,
the density of kinks is then very low
so that most of the time, there is no kink.
Step motion  then starts with the detachment
of an atom with $n=3$, at 
the rate $r_3$, followed by a fast kink-zipping process,
during which atoms at kink sites detach at a rate $r_2\gg r_3$. 
\footnote{The equilibrium concentration with
islands on top of the monolayer is $c_{eq}^{(2)}\sim {\rm e}^{-2J/T}$.
Thus, the attachment rate of atoms to
kinks is $\sim Dc_{eq}^{(2)}\sim {\rm e}^{-2J/T}\ll r_2$,
the kink atom detachment rate. Therefore,
backward kink motion is negligible.}
Since the step undergoes a total displacement $a$ at each zipping,
the velocity is equal to the total detachment rate
from a step segment of length $\lambda$
\begin{eqnarray}
V\sim \lambda r_3.
\label{e:V BT}
\end{eqnarray}

\begin{figure}
\centerline{\epsfysize=5cm\epsfbox{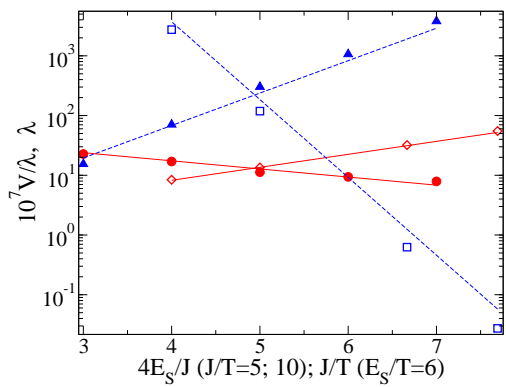}}
\caption{
Dewetting front velocity $V$ and width $\lambda$ of bilayer islands
in regime I, with IC2. 
Dashed lines correspond to Eq.(\ref{e:V BT}) with prefactor 0.15,
and solid lines to Eq.(\ref{e:lambda BT}) with prefactor 5.
Symbols are simulations. $\lambda$: for $J/T=10$ $\bullet$,
$E_S/T=6$ $\diamond$. $V/\lambda$:
for $J/T=10$ filled $\triangle$,
$E_S/T=6$ $\sqcap \! \! \! \! \sqcup$.  }
\label{fig:BT_step}
\end{figure}

Since $T/J$ is small,
islands grow in a irreversible fashion once
a dimer is formed on the second layer. 
The rate of formation of dimers
confined between two islands in a area $\sim \lambda^2$
is $\omega \sim D\lambda^2n_1^2$,
where $n_1$ is the adatom density, and $D=\nu_0a^2$ is the diffusion constant
on the second layer. Such a mean field expression of $\omega$
is valid up to logarithmic corrections \cite{politi}.
Adatoms are produced in bursts during
zipping processes, which last
during a time $t_z\sim \lambda/ar_2$, and
$n_1\sim r_2/D$.
The probability of dimer formation
during the zipping process is therefore:
$P_z=\omega t_z\sim\lambda^3r_2/aD$.
For self-consistency, 
islands must be formed 
after the step has moved over a distance $\lambda$,
so that $(\lambda/a)P_z\approx 1$,  leading to
\begin{eqnarray}
\lambda/a \sim r_2^{-1/4}\sim {\rm e}^{J/2T-E_S/4T}
\label{e:lambda BT}
\end{eqnarray}
As seen in Fig.\ref{fig:BT_step}, Eqs.(\ref{e:V BT})
and (\ref{e:lambda BT})
are in agreement with KMC simulations.

In regime II, temperature is higher, 
and mass exchange between mobile atoms and steps is fast,
so that mass transport is limited by
diffusion. Since the typical distance
between the edge of the monolayer and the edge of the
2D islands on top of it is  
$\lambda$,  one has
\begin{eqnarray}
V \sim {Da^2 \over \lambda}(c_{eq}^{(1)}-c_{eq}^{(2)})
\label{e:HT Vlambda}
\end{eqnarray}
where $c_{eq}^{(i)}$ is the concentration
of adatoms on top of the monolayer in equilibrium with
the atomic step separating $h=i-1$ and $h=i$.
From detailed balance,
$c_{eq}^{(1)}\sim {\rm e}^{(-2J+E_S)/T}$,
and $c_{eq}^{(2)}\sim {\rm e}^{-2J/T}$.

The separation $\lambda$ is determined by the standard 
nucleation theory, valid at high temperatures.
The Gibbs energy of a circular nucleus
of radius $R$ is the same for
islands or holes:
\begin{eqnarray}
G=2\pi R\gamma-\pi R^2 E_S/a^2 \, .
\label{e:G free}
\end{eqnarray}
where $\gamma$ is the line tension.
Since $\partial_RG<0$ for $R>R_{c}=a^2\gamma/E_S$,
nuclei with $R>R_c$ grow irreversibly. 
The island nucleation process
in the vicinity of the dewetting front
starts with a concentration
$c=c_{eq}^{(1)}$ in a typical area $\lambda^2$,
and ends with $c=c_{eq}^{(2)}$ in equilibrium
with the newly formed island of radius $R_{c}$. 
Mass conservation then 
imposes that 
$\lambda^2c_{eq}^{(1)}=\lambda^2c_{eq}^{(2)}+\pi R_{c}^2/a^2$,
so that:
\begin{eqnarray}
\lambda\approx  {\pi^{1/2}a\gamma \over 
E_S(c_{eq}^{(1)}-c_{eq}^{(2)})^{1/2}}
\label{e:lambda HT}
\end{eqnarray}
This relation is also valid for holes (we start from
$c_{eq}^{(2)}$, and end with $c_{eq}^{(1)}$, and
a hole of radius $R_c$).
The island-hole nucleation symmetry
is an important difference with the low temperature case.
In simulations on Fig.1 IIa and 1 IIb, we indeed observe
that both 2D islands and holes
form in the vicinity of the dewetting
front.

\begin{figure}
\centerline{\epsfysize=5cm \epsfbox{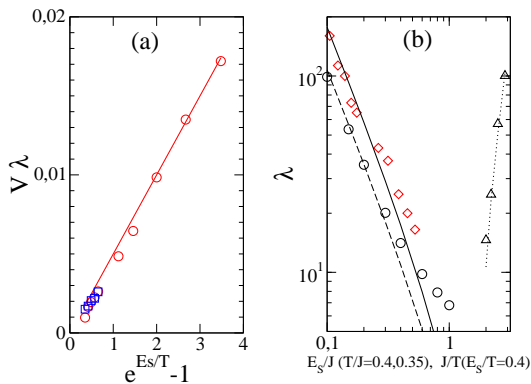}}
\caption{
Regime II: (a) $V\lambda$ is a linear function
to ${\rm e}^{E_S/T}-1$, at $J/T=2.86$, IC2. Squares are normal simulations
with $E_S$ small. Circles
are simulations where hole nucleation is forbidden.
(b) $\lambda$ for $T/J=0.35$, and IC2, varying $E_S/J$ ($\diamond$); 
$T/J=0.4$, and IC1, varying $E_S/J$ ($\circ$);
$E_S/T=0.4$, and IC2, varying $J/T$ ($\triangle$). 
Lines are from Eq.(\ref{e:lambda HT}) with prefactor $1$.
}
\label{fig:HT_step}
\end{figure}

The front velocity $V$ and the island 
width $\lambda$ obtained by simulations 
with IC2 agree well with theoretical expectation 
Eq. (\ref{e:HT Vlambda}) 
and (\ref{e:lambda HT}), as shown in Fig.3, at small $E_S$.
For large $E_S$, holes will be nucleated 
in the bulk of a monolayer and evaluation of
$V$ and $\lambda$ is hindered.
Therefore,
we have also performed simulations where the motion of atoms
with $n=4$ is forbidden. This prevents
the formation of holes in the film. Fig.\ref{fig:HT_step}
shows that simulation results 
with and without holes are in agreement
with Eqs.(\ref{e:HT Vlambda}) and (\ref{e:lambda HT}).
For the calculation of $\lambda$ , we have used the
expression of $\gamma$ from the Ising model \cite{Rottman}.

Having a fairly complete description of the 
motion of the dewetting front in regimes I and II,
we shall now proceed with the analysis
of nucleation and growth of Dewetted Zones (DZ)
by means of a simple
geometric model. In this model, DZs 
are nucleated
with the rate ${\cal J}$ and expand isotropically in the monolayer.
The periphery of a DZ is initially
a circle expanding with the constant  
velocity $V$,
which was determined above.
While they grow, DZs
will meet. The sum ${\cal L}$
of the perimeters of the 
DZs if they did not see each other (i.e. the sum of the lengths
of the circles) obeys
\begin{eqnarray}
\partial_t{\cal L}=2\pi VN, 
\label{e:evol_ell}
\end{eqnarray}
where $N$ is the number of DZs.
Moreover, new DZs can only form outside pre-existing
DZs, so that
\begin{eqnarray}
\partial_tN={\cal J}({\cal A}-{\cal A}_{DZ}),
\label{e:evol_n}
\end{eqnarray}
where ${\cal A}$ is the total area,
and ${\cal A}_{DZ}$ is the total area of the DZs.
Because of the overlap
of the growing DZs, the length 
${\cal L}_{DZ}$ of the  frontier between
the DZs and the monolayer is smaller than ${\cal L}$.
In a mean field picture, we assume that
${\cal L}_{DZ}\approx {\cal L}(1-{\cal A}_{DZ}/{\cal A})$,
so that
\begin{eqnarray}
\partial_t{\cal A}_{DZ}={\cal L}_{DZ}V
\approx {\cal L}(1-{\cal A}_{DZ}/{\cal A})V\, .
\label{e:evol_theta}
\end{eqnarray}
Instead of analyzing the full solution
of (\ref{e:evol_ell},\ref{e:evol_n},\ref{e:evol_theta}),
we shall focus on the behavior of 
the uncoverage
$\theta=(1/2){\cal A}_{DZ}/{\cal A}$
(the fraction of the substrate surface which is not covered).
At short times,
one finds: 
\begin{eqnarray}
\theta=\pi V^2{\cal J}t^3/6,
\end{eqnarray}
and at long times $\theta\rightarrow 1/2$.
The saturation time $t_{sat}$, which separates these
two regimes, may be defined 
as the time where $|\partial_{tt}\theta|$
is maximum. 
We find:
\begin{eqnarray}
t_{sat}=\alpha V^{-2/3}{\cal J}^{-1/3}
\label{e:tc}
\end{eqnarray}
where $\alpha\approx 0.85$.
The time $t_{sat}$ may be interpreted as the
time needed for the dewetting process
to invade the whole monolayer.

\begin{figure}
\centerline{\epsfysize=4cm\epsfbox{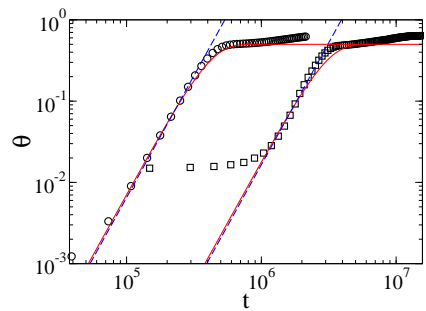}}
\caption{
Uncoverage $\theta$ as a function
of time (in Monte Carlo Steps per Site).
Simulations with: $J/T=5$, $E_S/J=1$, $\sqcap \! \! \! \! \sqcup$,
and $J/T=2.5$, $E_S/J=0.15$ $\circ$.
Dashed lines are $\sim t^{3}$, and  solid lines are the
solution of the geometrical model. 
}
\label{fig:scaling theta}
\end{figure}

As shown on Fig.\ref{fig:scaling theta},
the $t^3$-scaling behavior of $\theta(t)$ is in agreement
with the simulations, except for very short times,
where $\theta$ originates
from the thermal roughening of the monolayer.
In the following,
we shall see how $t_{sat}$ from the simulations
compares with Eq.(\ref{e:tc}).
To do so, we need to derive
the expression of ${\cal J}$.

In regime I, we resort to
rate equations to describe the nucleation process.
A hole of size N is denoted as a N-hole.
The initial  1-hole concentration $\rho_1$
essentially results from adatom-pair
formation, at rate $a^{-2}r_4$, and annihilation,
at the rate $Dn_1\rho_1$, so that
\begin{eqnarray}
\rho_1\approx n_1\approx (a^{-2}r_4/D)^{1/2}=a^{-2}{\rm e}^{(-2J+E_S/2)/T}.
\label{e:rho1 BT}
\end{eqnarray}
Let us denote $\sigma_{N,N\pm1}$ the rate at which 
a $N$-hole transforms into a ($N\pm 1$)-hole.
The rates of size-decreasing transitions are related to the
attachment of a mobile atom:
$\sigma_{2,1}\sim\sigma_{3,2}\sim\sigma_{4,3}\sim Dn_1$.
Other rates results from the detachment of a 3-neighbors
atom: $\sigma_{1,2}\sim\sigma_{2,3}\sim\sigma_{4,5}\sim r_3$,
or 2-neighbors atoms: $\sigma_{3,4}\sim\sigma_{5,6}\sim r_2$.
We only consider compact square 4-holes,
because they are the most stable 4-holes \cite{unpublished}.
As a consequence, we also only consider $L$-shaped
3-holes, which are the ones leading to
square 4-holes. Assuming a steady-state, 
and following the standard Becker-D\"oring analysis, 
we obtain the hole formation rate:
\begin{eqnarray}
{\cal J} =\rho_1
\left({1 \over \sigma_{12}}
+{\sigma_{21} \over \sigma_{12}\sigma_{23}}
+{\sigma_{21}\sigma_{32} \over \sigma_{12}\sigma_{23}\sigma_{34}}
+....\right)^{-1},
\end{eqnarray}
which leads to
\begin{eqnarray}
{\cal J} \approx \rho_1 {\sigma_{12}\sigma_{23} \over \sigma_{21}}
\sim {\rm e}^{(-6J+2E_S)/T},
\label{e:j4}
\end{eqnarray}
when $J<E_S<2J$. This regime correspond
to a critical size of $4$.
Using Eqs.(\ref{e:V BT},\ref{e:lambda BT},\ref{e:j4}) 
in Eq.(\ref{e:tc}) leads to:
\begin{eqnarray}
t_{sat}\sim \nu_0^{-1}{\rm e}^{(11J/3-7E_S/6)/T}.
\label{e:tc BT}
\end{eqnarray}
Fig.5a shows the agreement of Eq.(\ref{e:tc BT}) with the simulations.
When $E_S<J$, the critical size will be larger than 4, leading to
deviations from (\ref{e:tc BT})\cite{unpublished}.

\begin{figure}
\centerline{\epsfysize=5cm\epsfbox{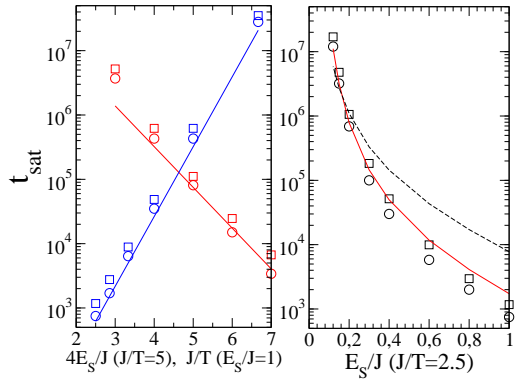}}
\caption{
$t_{sat}$ ($\circ$), and $t_{1/2}$ ($\sqcap \! \! \! \! \sqcup$)
--the time at which $\theta$ reaches $1/2$--
as a function of $E_S/J$ and $J/T$, with IC1.
(a) Low temperatures, varying $J/T$ with $E_S/J=1$, and 
varying $E_S/J$, with $J/T=5$.
The solid lines correspond to Eq.(\ref{e:tc BT})
with a prefactor of $1.5$
(b) High temperatures, varying $E_S/J$, with $J/T=2.5$.
The solid line
corresponds to Eq.(\ref{e:tc HT}) with a prefactor
of $25$, and $\gamma_{eff}$, while
the dashed line corresponds to Eq.(\ref{e:tc HT})
with $\gamma$.
}
\label{fig:t1/2 HT BT}
\end{figure}

In regime II, the thermodynamic
theory of nucleation applies \cite{Saito},
and the formation rate of holes reads:
\begin{eqnarray}
{\cal J}=\rho_0\Gamma_{+c}\left({G_c'' \over 2\pi T}\right)^{1/2}
{\rm e}^{-G_c/T}\, ,
\label{e:j nucl theory}
\end{eqnarray}
where $G_c$ is the value of $G$
at the critical island size $s_c$,
and $G_c''=\partial_{ss}G|_{s=s_c}$,
with $s=\pi R^2$,
$\Gamma_{+c}$ is the rate at which 
atoms attach to the hole when $R=R_c$,
and $\rho_0\approx 1/a^2$ is the monolayer density.
The concentration in the
vicinity of the holes at the critical size
should be at equilibrium,
i.e. $c=c_{eq}^{(2)}\sim {\rm e}^{-2J/T}$, 
so that $\Gamma_{+c}=2\pi R_cc_{eq}^{(2)}(D/a)$.
Thus, Eq.(\ref{e:j nucl theory}) leads to:
\begin{eqnarray}
{\cal J}={D \over  a^{4}}\, 
E_S^{1/2}  T^{-1/2} {\rm e}^{(-2J-\pi\gamma^2a^2/E_S)/T}.
\end{eqnarray}
Upon substitution in Eq(\ref{e:tc}), and using 
(\ref{e:HT Vlambda},\ref{e:lambda HT}) we find:
\begin{eqnarray}
t_{sat} \sim D^{-1}
{a^{-4/3}\gamma^{2/3}T^{1/6} \over E_S^{5/6}({\rm e}^{E_S/T}-1)}
{\rm e}^{(8J+\pi\gamma^2a^2/E_S)/3T}.
\label{e:tc HT}
\end{eqnarray}
We have performed simulations
at $J/T=2.5$. From the Ising model\cite{Rottman},
we find $\gamma/J\approx 0.27$.
The line tension $\gamma$ at such high
temperatures must be corrected in order
to account for the non-ideal
character of the 2D gas of adatoms around the cluster.
Using the results of Ref.\cite{sethna},
we find: $\gamma_{eff}/J \approx 0.42$
\footnote{The results of Ref.\cite{sethna} are obtained
in a strictly 2D model.
The thermal roughness of the layer, which is reinforced
when $E_S\neq 0$, is neglected here.
Following Ref.\cite{sethna}, 
one essentially needs to add 
a multiplicative factor to $\gamma$.
Therefore, this should not  
strongly affect the temperature dependence of Eq.(\ref{e:lambda HT}).}.
The high temperature limit is obtained
when $R_c\gg 1$, i.e. when $E_S/J\ll \pi^{1/2}a^2\gamma_{eff}/J\approx 0.74$.
The simulation results with $0.12\leq E_S/J \leq 1$ 
reported on Fig.\ref{fig:t1/2 HT BT} indeed agree with
Eq.(\ref{e:tc HT}) for small $E_S$.

Orders of magnitude may be obtained
for the case of Au/graphite.
Using $J\approx 0.5eV$\cite{Bahn2001}, 
$E_S\approx 0.1 eV$\cite{Smith2006},
and a diffusion barrier $\sim 0.24 eV$\cite{Anton1998},
we find that we are in regime II, 
with $t_{sat}\sim 10$ min, and $\lambda\sim 5\times 10^3$a
at $800K$. 
Moreover, the dewetting of a nanometer-thick
Ag/Si(111) \cite{Thuermer}, lead to patterns
similar to those obtained here in regime II,
suggesting that our finding may be relevant
for films with a small number of atomic layers.
Quantitative experimental analysis
of the dewetting process in various systems
is still lacking though.

We have focused on the formation of the
bilayer islands because this process determines
the break-up time $t_{sat}$, and
the initial lengthscale $\lambda$.
The subsequent dynamics combines
processes which have been separately studied
in the literature:  layer-by-layer
thickening of the islands\cite{tosh}, coupled to
their sintering\cite{evans-liu} and Ostwald ripening.
We hope to report along these lines in the future.

We acknowledge support from nanomorphog\'en\`ese
ANR-PNANO grant, and from the computing
facilities at CIMENT, Grenoble. YS acknowledges support from 
JSPS.

\end{document}